\documentclass[10pt,journal,epsfig]{IEEEtran}
\usepackage{graphicx}
\usepackage{subfigure,color}
\usepackage{amssymb}
\usepackage{cite,tcolorbox}
\usepackage{amsmath,setspace}
\usepackage{bm,comment,epstopdf}
\usepackage{algorithm}
\usepackage{algpseudocode}
\makeatletter

\newcommand{\Rmnum}[1]{\expandafter\@slowromancap\romannumeral #1@}
\newcommand{\mv}[1]{\mbox{\boldmath{$ #1 $}}}

\makeatother

\newtheorem{lemma}{Lemma}

\begin{document}
\title{\LARGE{Movable-Antenna Position Optimization: A Graph-based Approach}}
\author{Weidong Mei, \IEEEmembership{Member, IEEE}, Xin Wei, Boyu Ning, \IEEEmembership{Member, IEEE}, Zhi Chen, \IEEEmembership{Senior Member, IEEE}, and Rui Zhang, \IEEEmembership{Fellow, IEEE}
\thanks{W. Mei, X. Wei, B. Ning, and Z. Chen are with the National Key Laboratory of Wireless Communications, University of Electronic Science and Technology of China, Chengdu 611731, China (e-mail: wmei@uestc.edu.cn, xinwei@std.uestc.edu.cn, boydning@outlook.com, chenzhi@uestc.edu.cn).}
\thanks{R. Zhang is with School of Science and Engineering, Shenzhen Research Institute of Big Data, The Chinese University of Hong Kong, Shenzhen, Guangdong 518172, China (e-mail: rzhang@cuhk.edu.cn). He is also with the Department of Electrical and Computer Engineering, National University of Singapore, Singapore 117583 (e-mail: elezhang@nus.edu.sg).}}
\maketitle

\begin{abstract}
Fluid antennas (FAs) and movable antennas (MAs) have emerged as promising technologies in wireless communications, which offer the flexibility to improve channel conditions by adjusting transmit/receive antenna positions within a spatial region. In this letter, we focus on an MA-enhanced multiple-input single-output (MISO) communication system, aiming to optimize the positions of multiple transmit MAs to maximize the received signal power. Unlike the prior works on continuously searching for the optimal MA positions, we propose to sample the transmit region into discrete points, such that the continuous antenna position optimization problem is transformed to a discrete sampling point selection problem based on the point-wise channel information. However, such a point selection problem is combinatory and challenging to be optimally solved. To tackle this challenge, we ingeniously recast it as an equivalent fixed-hop shortest path problem in graph theory and propose a customized algorithm to solve it {\it \textbf{optimally}} in polynomial time. To further reduce the complexity, a linear-time sequential update algorithm is also proposed to obtain a high-quality suboptimal solution. Numerical results demonstrate that the proposed algorithms can yield considerable performance gains over the conventional fixed-position antennas with/without antenna selection.
\end{abstract}

\section{Introduction}
Fluid antenna (FA) and movable antenna (MA) technologies, traditionally explored in the realm of antenna design, have recently drawn great attention in wireless communications, owing to their ability to dynamically adjust the positions of transmit/receive antennas within a given region\cite{zhu2024movable,wong2020fluid}. As compared to the conventional fixed-position antennas (FPAs) which may experience deep fading at specific positions for a given time and/or frequency resource block, FAs and MAs offer the advantage of flexible antenna movement to circumvent such deep-fading positions, thereby reshaping the wireless channels into a more favorable condition for communication performance enhancement\cite{wong2023fluid,zhu2024modeling}.

Motivated by the promising benefits of the FA/MA technology, prior studies have delved into the antenna position optimization under various system setups. Notably, the intricate coupling between channel responses and antenna positions gave rise to a new challenge in finding optimal antenna positions to maximize communication rate. To tackle this challenge, the authors in \cite{zhu2024modeling,ma2023mimo,qin2024antenna,zhu2024multiuser,hu2024fluid,hu2024secure} proposed a field-response-based channel model in the angular domain to characterize the end-to-end channel, which appears to be a highly nonlinear function with respect to (w.r.t.) the antenna positions. As such, a variety of iterative algorithms, e.g., alternating optimization (AO)\cite{ma2023mimo}, successive convex approximation (SCA)\cite{qin2024antenna}, gradient descent\cite{zhu2024multiuser,hu2024fluid,hu2024secure}, among others, were proposed to search for the locally optimal antenna positions in a continuous space. In contrast, an alternative approach is by discretizing the transmit or receive region into a multitude of ports/sampling points for moving FAs/MAs over them. Based on the channel state information (CSI) for each port or sampling point, the antenna positions can be optimized by solving a port/point selection problem\cite{chai2022port,new2023fluid,wu2023movable} subject to the constraints on minimum antenna spacing to avoid mutual coupling. As compared to the continuous optimization of antenna positions, such a discrete port/point selection method is practically easier to implement, but may incur exorbitant computational complexity for finding the optimal solutions due to the generally required high level of discretization to reduce performance loss as compared to continuous position optimization. In the case of a single FA/MA, the optimal port selection was obtained via a one-dimensional search in \cite{chai2022port,new2023fluid}. However, in the case of multiple FAs/MAs, the port/point selection problem becomes more challenging to be optimally solved. In \cite{wu2023movable}, the authors introduced a generalized Bender's decomposition method to obtain the optimal point selection in an MA-enhanced multi-user system. However, this method entails an exponential complexity order in the worst case, which may not be affordable in practice. 

\begin{figure}[!t]
\centering
\includegraphics[width=3in]{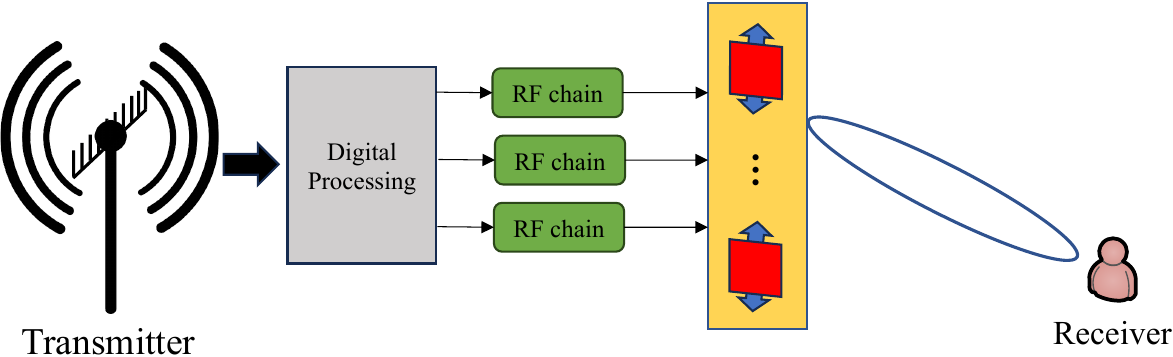}
\DeclareGraphicsExtensions.
\caption{MA-enhanced MISO communication system.}\label{sysmod}
\vspace{-9pt}
\end{figure}
In this letter, we propose an efficient graph-based algorithm to solve the port/point selection problem {\it optimally} for an MA-enhanced multiple-input single-output (MISO) communication system as shown in Fig.\,\ref{sysmod}, with multiple MAs equipped in a confined region at the transmitter (e.g., an access point shown in Fig.\,\ref{sysmod}). In particular, with the optimal maximum-ratio transmission (MRT), we aim to optimize the positions of all MAs over a given set of discrete points to maximize the received signal power at the receiver with a single FPA. However, this point selection optimization problem is combinatory and thus difficult to be optimally solved. To tackle this challenging problem, we ingeniously model the discrete points as vertices in a graph and define its edge weights based on the CSI, thereby recasting the point selection problem as an equivalent {\it fixed-hop} shortest path problem in graph theory, which is then optimally solved by a customized polynomial-time algorithm. To further reduce the computational complexity, a linear-time suboptimal sequential update algorithm is also proposed by sequentially selecting the points for MAs. Numerical results show that the proposed algorithms significantly outperform the conventional FPAs with and without antenna selection.

{\it Notations:} Bold symbols in capital letter and small letter denote matrices and vectors, respectively. The conjugate transpose of a vector or matrix is denoted as ${(\cdot)}^{H}$. ${\mathbb{R}}^n$ (${\mathbb{C}}^n$) denotes the set of real (complex) vectors of length $n$. $\lVert \mv a \rVert$ denotes the Euclidean norm of the vector $\mv a$. For a complex number $s$, $s \sim \mathcal{CN}(0,\sigma^2)$ means that it is a circularly symmetric complex Gaussian (CSCG) random variable with zero mean and variance $\sigma^2$. For two sets $A$ and $B$, $A \cup B$ and $A \cap B$ denote their union and intersection, respectively. $\lvert A \rvert$ denotes the cardinality of the set $A$. ${n \choose k} = \frac{n!}{k!(n-k)!}$ denotes the number of ways to choose $k$ elements from a set of $n$ elements. ${\cal O}(\cdot)$ denotes the order of complexity. \vspace{-6pt}

\begingroup
\allowdisplaybreaks
\section{System Model and Problem Formulation}
\subsection{System Model}
As shown in Fig.\,\ref{sysmod}, we consider a MISO communication system, where the transmitter is equipped with $N$ MAs, while the receiver is equipped with a single FPA. In particular, we consider a linear transmit MA array of length $L$, over which the positions of the $N$ MAs can be flexibly adjusted. Note that the proposed algorithms can be extended to a two-dimensional (2D) transmit array. Let ${\cal N}=\{1,2,\cdots,N\}$ denote the set of all MAs. We assume that the channel between the transmitter and receiver is quasi-static (i.e., slowly-varying), such that the MAs in ${\cal N}$ can be moved to their respective optimized positions with negligible time as compared to the channel coherence time.

For the ease of implementing antenna movement, we uniformly sample the transmit MA array into $M\;(M \gg N)$ discrete positions, with an equal spacing between any two adjacent sampling points given by $\delta_s=L/M$. Hence, the position of the $m$-th sampling point is given by $s_m=\frac{mL}{M}, m \in {\cal M}=\{1,2,\cdots,M\}$, with ${\cal M}$ denoting the set of all sampling points within the MA array. As such, the position of each MA can be selected as one of the sampling points in ${\cal M}$. Let $a_n, a_n \in \cal M$ denote the index of the selected sampling point for the $n$-th MA. Thus, the position of the $n$-th MA can be expressed as $s_{a_n}=\frac{a_nL}{M}, n \in \cal N$. To avoid the mutual coupling between MAs with a finite size, we consider a minimum distance, $d_{\min}$, between any pair of MAs. Thus, it must hold that
\begin{equation}\label{dmin}
\lvert a_i - a_j \rvert \ge a_{\min}, \forall i,j \in {\cal N}, i \ne j,
\end{equation} 
where $a_{\min}\!=\! d_{\min}/\delta_s \gg 1$, which is assumed to be an integer. It follows that the MA position optimization is equivalent to the sampling point selection from $\cal M$ subject to (\ref{dmin}).

Denote by $h_m \in {\mathbb{C}}, m \in \cal M$ the baseband-equivalent channel from the $m$-th sampling point to the receiver. To investigate the performance limit of the proposed algorithm, we assume that $h_m$'s are known at the transmitter, while in practice they can be acquired by applying the compressed sensing (CS)-based channel estimation\cite{ma2023compressed}, the sparse channel reconstruction\cite{xu2023channel} or the spatial interpolation based on channel measurements at a subset of the sampling points\cite{zhang2023successive}. Based on the above, the channel from the transmitter to the receiver is expressed as \[{\mv h}(\{a_n\})=[h_{a_1},h_{a_2},\cdots,h_{a_N}]^H.\]

Let ${\mv w}_t \in {\mathbb C}^N$ and $P_t$ denote the transmit beamforming vector and transmit power, respectively. For any given indices of the selected sampling points $a_n, n \in \cal N$, the transmitter should apply the MRT to maximize the received signal power, i.e., ${\mv w}_t = \sqrt{P_t}{\mv h}(\{a_n\})/\lVert {\mv h}(\{a_n\}) \rVert$. The resulting maximum received power is given by
\begin{equation}\label{recPw}
P_r(\{a_n\}) = \lvert {\mv w}_t^H{\mv h}(\{a_n\}) \rvert^2=P_t\sum\limits_{n=1}^N{\lvert h_{a_n} \rvert^2}.
\end{equation}\vspace{-15pt}

\subsection{Problem Formulation}
In this letter, we aim to maximize the received signal power in (\ref{recPw}) by optimizing the MA positions over the discrete sampling points, i.e., $\{a_n\}_{n \in \cal N}$. The corresponding optimization problem is formulated as
\begin{align}
{\text{(P1)}}\; \mathop {\max}\limits_{\{a_n\}}&\; \sum\limits_{n=1}^N{\lvert h_{a_n} \rvert^2} \nonumber\\
\text{s.t.}\;\;& a_n \in {\cal M}, \forall n \in {\cal N},\label{op2a}\\
&\lvert a_i - a_j \rvert \ge a_{\min}, \forall i,j \in {\cal N}, i \ne j, \label{op2b}
\end{align}
where the constant $P_t$ is omitted in the objective function of (P1). It is worth noting that as compared to the continuous MA position optimization problems under the field-response based channel model (see e.g., \cite{ma2023mimo,qin2024antenna,zhu2024multiuser,hu2024fluid,hu2024secure}), (P1) is a discrete optimization problem by avoiding the highly nonlinear channels w.r.t. the MA positions. Moreover, (P1) is also different from the conventional antenna selection problem due to the minimum distance constraints in (\ref{op2b}) and the much larger number of sampling points ($M$) as compared to the number of antennas selected for transmission in antenna selection. 

However, (P1) is challenging to be optimally solved in general due to its combinatorial nature and the constraints in (\ref{op2b}). One straightforward approach to optimally solve (P1) is by enumerating all possible MA positions. However, it can be shown that this incurs an exorbitant complexity in the order of $M-(a_{\min}-1)(N-1) \choose N$, which may not be applicable to a large size of antenna array with large $M$ and/or $N$ values in practice. It is also noted that in the special case of $\delta_s=d_{\min}$, we have $M=L/d_{\min}$ and $a_{\min}=1$, and thus the constraints in (\ref{op2b}) can be removed. As a result, the above complexity order reduces to $M \choose N$, which is the same as that of conventional antenna selection (i.e., selecting $N$ out of $M$ FPAs).\vspace{-9pt}

\section{Proposed Algorithms for (P1)}\label{alg}
To solve (P1), we propose in this section an efficient graph-based optimal solution and a lower-complexity suboptimal solution, respectively.\vspace{-6pt}

\subsection{Optimal Solution by Graph Optimization}
First, we show that (P1) can be recast as a shortest path problem (SPP) in graph theory subject to (\ref{op2b}). Specifically, we construct a directed weighted graph $G=(V,E)$. The vertex set $V$ is given by the set of all sampling points, i.e., $V={\cal M}$. Without loss of optimality, we consider that the MA indices are selected in order from one sampling point $i$ to a farther point $j$ from the reference position 0.\footnote{Note that in the case of a 2D transmit array, the sampling points can be ordered based on their column indices initially, followed by their row indices.} Accordingly, the edge set $E$ is defined as
\begin{equation}\label{edgeset}
E=\{(i,j)|j-i \ge a_{\min}, i, j \in {\cal M}\},
\end{equation}
i.e., an edge exists from vertex $i$ to vertex $j$ if and only if their corresponding sampling points are sufficiently separated. It is not difficult to verify that $\lvert E \rvert=\frac{(M-a_{\min})(M-a_{\min}+1)}{2}$. Furthermore, we set the weight of each edge $(i,j)$ in $E$ as $W_{i,j}=\lvert h_i \rvert^2$. Based on the above, it can be easily shown that the objective value of (P1) by any MA indices $a_n, n \in {\cal N}$ is equal to the sum of edge weights of the path $a_1 \rightarrow a_2 \rightarrow \cdots \rightarrow a_N$ in $G$.

Next, to facilitate our SPP formulation, we add two ``dummy'' vertices 0 and $M+1$ to $G$, and the vertex set becomes $\tilde V = V \cup \{0,M+1\}$. Moreover, we add an edge from vertex 0 to each vertex in $V$ and from each vertex in $V$ to vertex $M+1$. As such, the edge set $E$ in (\ref{edgeset}) becomes
\begin{equation}
\tilde E=E \cup \{(0,j)|j \in V\} \cup \{(j,M+1)|j \in V\},
\end{equation}
and $\lvert \tilde E \rvert=\lvert E \rvert + 2M$. Finally, for the new graph $\tilde G=(\tilde V,\tilde E)$, we set the weights of its edges as
\begin{equation}
\tilde W_{i,j}=
\begin{cases}
	0, &{\text{if}}\;i=0\\\
	-W_{i,j}, &{\text{otherwise}}
\end{cases}, \forall (i,j) \in {\tilde E}.
\end{equation}
It follows that the objective value of (P1) by any MA indices $a_n, n \in {\cal N}$ is equal to the negative sum of edge weights of the path $0 \rightarrow a_1 \rightarrow a_2 \rightarrow \cdots \rightarrow a_N \rightarrow M+1$. Hence, problem (P1) is equivalent to the following SPP, (P2).
\begin{tcolorbox}[standard jigsaw, opacityback=0]
  (P2)\;{\it Find the $(N+1)$-hop shortest path (i.e., with the minimum sum of the weights of the constituent edges) from vertex 0 to vertex $M+1$ in $\tilde G$.}
\end{tcolorbox}

It is worth noting that different from the general SPP, (P2) specifies the hop number of the shortest path. Nonetheless, as $\tilde G$ is a directed acyclic graph, (P2) can still be optimally solved in polynomial time, as shown in the following lemma.
\begin{lemma}[\!\!\cite{cheng2004finding}]\label{dp}
	Denote by $p^k(0,i)$ the $k$-hop shortest path from vertex 0 to another vertex $i, i \in \tilde V$. Then, if $k=0$, we have $p^k(0,i)=(0,i)$. If $(0,i)$ does not exist, we assume that there exists a virtual edge from vertex 0 to vertex $i$ with an infinite weight. For $k \ge 1$, let ${\cal N}_i=\{j|(j,i) \in \tilde E\}$ denote the set of all incoming neighbors of vertex $i$. Then, $p^k(0,i)$ is the shortest path among the paths $p^{k-1}(0,x)+(x,i), x \in {\cal N}_i$, i.e., the paths obtained by concatenating $p^{k-1}(0,x)$ with the edge $(x,i)$ for all $x \in {\cal N}_i$. If there exists no $k$-hop path from vertex 0 to vertex $i$, we assume that there exists a virtual $k$-hop path with an infinite weight.
\end{lemma}

Based on Lemma \ref{dp}, we can obtain the optimal solution to (P2) as $p^{N+1}(0,M+1)$ recursively by applying the dynamic programming (DP). The overall complexity of the DP procedures is given by ${\cal O}(N\lvert \tilde E \rvert)={\cal O}(NM^2)$\cite{cheng2004finding}.\vspace{-8pt}

\subsection{Suboptimal Solution by Sequential Update}
Although the optimal solution to (P1) can be obtained by the graph-based algorithm, its worst-case complexity is quadratic in the number of sampling points $M$, which can still be high with increased $M$ for improving the sampling resolution. To address this issue, we propose a lower-complexity sequential update algorithm to solve (P1) sub-optimally, by sequentially selecting the sampling points for MAs.

Specifically, given an initial selection of the sampling points denoted as $\{a_{n,\text{ini}}, n \in {\cal N}\}$, we consider that the $n$-th initial sampling point, i.e., $a_{n,\text{ini}}$, needs to be updated in the $n$-th iteration of the sequential search, with the $i$-th initial sampling point ($1 \le i \le n-1$) updated as $a_i^*$ already. As such, in the $n$-th iteration, $a_{n,\text{ini}}$ can only be selected from the set
\begin{align}
	\Psi_n\!=&\{m|m \in {\cal M}, \lvert m-a^*_i \vert \ge a_{\min}, 1 \!\le\! i \!\le\! n-1, \lvert m-a_{j,\text{ini}} \vert\nonumber\\
	& \ge a_{\min}, n+1 \le j \le N, \}, 2 \le n \le N-1,\label{psi}
\end{align}
and we set $\Psi_1=\{m|m \in {\cal M}, \lvert m-a_{j,\text{ini}} \vert \ge a_{\min}, 2 \le j \le N\}$ and $\Psi_N=\{m|m \in {\cal M}, \lvert m-a^*_i \vert \ge a_{\min}, 1 \le i \le N-1\}$. To maximize the objective function of (P1), $a_{n,\text{ini}}$ should be updated as
\begin{equation}\label{an}
	a^*_n=\arg\mathop {\max}\limits_{m \in \Psi_n}\lvert h_m \rvert^2.
\end{equation}
Next, the update of the $(n+1)$-th sampling point follows by updating $\Psi_{n+1}$ based on (\ref{psi}). It can be readily shown that $M-(2a_{\min}-1)(N-1) \le \lvert \Psi_n \rvert \le M$. Hence, the complexity of our proposed sequential update algorithm is in the order between $MN-N(N-1)(2a_{\min}-1)$ and $MN$, which is linear in $M$ for any given $N$ and thus lower than that of the graph-based optimal algorithm. However, it should be mentioned that the sequential update algorithm may only achieve suboptimal performance, since the sets $\Psi_n, n \in \cal N$ depend on the initial sampling point selection and the order of the selected sampling points, and thus some optimal sampling points may be eliminated in the update of these sets, as will be shown in Section \ref{sim} via simulation. The main steps of the sequential update algorithm are summarized in Algorithm \ref{Alg}. 
\begin{algorithm}
  \caption{Sequential Update Algorithm for Solving (P1)}\label{Alg}
  \begin{algorithmic}[1]
    \State Initialize $n=1$, $a_{n,\text{ini}}, n \in {\cal N}$, and $\Psi_1$.
    \While {$n \le N$}
    \State Determine $a_n^*$ based on (\ref{an}) and update $a_{n,\text{ini}}=a_n^*$.
    \State Determine $\Psi_{n+1}$ based on (\ref{psi}).
    \State Update $n=n+1$.
    \EndWhile
    \State Output $\{s_{a_n^*}\}$ as the optimized MA positions.
    \end{algorithmic}
\end{algorithm}
\vspace{-3pt}
\section{Numerical Results}\label{sim}
In this section, we provide numerical results to evaluate the performance of our proposed optimal graph-based algorithm and the suboptimal sequential update algorithm. Unless otherwise stated, the simulation settings are as follows. The carrier frequency is 5 GHz, and thus the wavelength is $\lambda=0.06$ meter (m). The number of transmit MAs is $N = 8$, while the length of the linear transmit array is $L=0.36\;{\text{m}}=6\lambda$. The minimum distance between any two MAs is set to $d_{\min} = \lambda/2$. Let $D$ and $\alpha$ denote the distance from the transmitter to the receiver and the path-loss exponent, respectively, which are set to $D=100$ m and $\alpha=2.8$. To generate the channels for the sampling points, i.e., $h_m, m \in \cal M$, we consider the field-response channel model in \cite{zhu2024modeling,ma2023mimo,qin2024antenna,zhu2024multiuser,hu2024fluid,hu2024secure}, with the number of transmit paths set to 9. Let $\gamma_i$ denote the channel response coefficient for the $i$-th transmit path, which is assumed to follow CSCG distribution with $\gamma_i \sim {\cal{CN}}(0,\beta D^{-\alpha}l_i)$, where $\beta$ denotes the path loss at the reference distance of 1 m, and $l_i$ denotes the ratio of the average power gain of the $i$-th transmit path to that of all transmit paths. We set $\beta=-46$ dB, while the values of $l_i$'s are randomly generated and then normalized to satisfy $\sum\nolimits_{i}l_i=1$. The angle of departure (AoD) from the transmit array for each transmit path is assumed to be a uniformly distributed variable over $[0,\pi]$. The transmit signal-to-noise ratio (SNR) is $P_t/\sigma^2=100$ dB, with $\sigma^2$ denoting the receiver noise power. 

We compare our proposed algorithms for MAs with the following two benchmarks, namely, FPAs without (w/o) or with (w/) antenna selection (AS). In the first benchmark (FPA w/o AS), we assume that $N$ FPAs are deployed symmetrically to the center of the transmit array and separated by the minimum distance $d_{\min}$. In the second benchmark (FPA w/ AS), $L/d_{\min}=12$ FPAs are deployed along the entire transmit linear array and separated by the minimum distance $d_{\min}$. Among them, $N$ antennas with the largest channel power gains with the receiver are selected to maximize the received SNR. Note that this scheme is equivalent to our proposed MA scheme when $\delta_s=d_{\min}$ and thus $M=12$. In the sequential update algorithm, the sampling point selection is initialized as that by FPA w/ AS. All the results are averaged over 1000 independent channel realizations.

\begin{figure}[!t]
\centering
\includegraphics[width=2.5in]{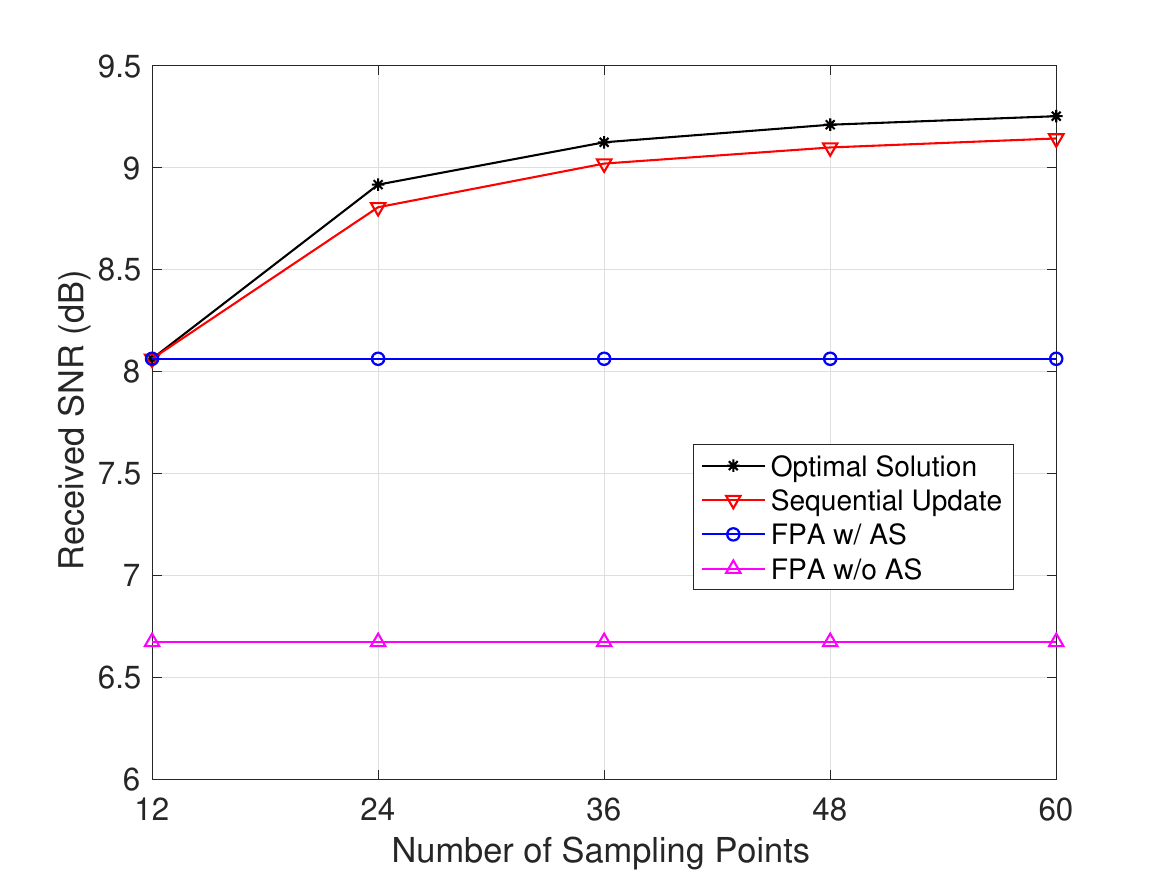}
\DeclareGraphicsExtensions.
\caption{Received SNR versus the number of sampling points.}\label{SNRvsNumSamp}
\vspace{-12pt}
\end{figure}
First, we plot the received SNR versus the number of sampling points $M$ in Fig.\,\ref{SNRvsNumSamp}. It is observed that the performance of the proposed algorithms improves with $M$ thanks to the refined sampling resolution, as expected, but the growth rate becomes slower with increasing $M$. Particularly, when $M \ge 48$, further increasing $M$ can barely improve the received SNR, implying that a sufficiently high resolution has been achieved. This also suggests that a moderate number of sampling points suffices to achieve near-optimal performance of continuous searching. It is also observed that the suboptimal sequential update algorithm can yield comparable performance to the optimal graph-based algorithm, and both algorithms can achieve approximately 1.1 dB and 2.5 dB higher received SNR than the FPA w/ AS and FPA w/o AS benchmark schemes, respectively, when $M$ is sufficiently large.

\begin{figure}[!t]
\centering
\includegraphics[width=2.5in]{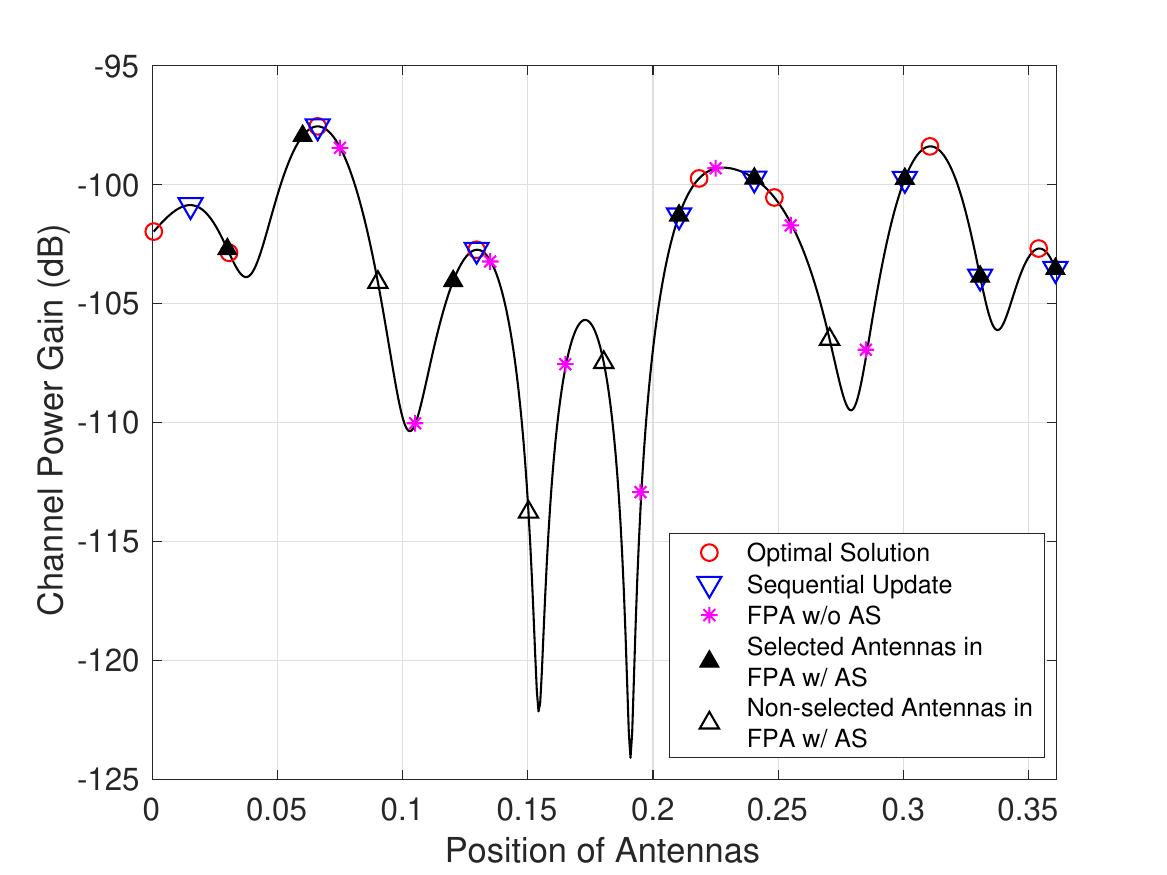}
\DeclareGraphicsExtensions.
\caption{Optimized antenna positions in different schemes.}\label{MA_Position}
\vspace{-12pt}
\end{figure}
To gain more insights into the antenna position optimization, we plot in  Fig.\,\ref{MA_Position} the channel power gains over the positions on the entire transmit array and mark the positions of the transmitting antennas of the MA scheme by our proposed algorithms with $M=48$ and the two benchmark schemes. It is observed that there exist multiple local maximum and minimum channel-gain positions along the transmit array, and their difference can exceed 25 dB. Some local maximum channel-gain positions are selected as the optimized MA positions, while other local maximum channel-gain positions, e.g., those at 0.015 m and 0.17 m, are not selected. Instead, two positions near each of them are selected as optimal MA positions. Moreover, it is observed that some positions near the local channel-gain minimums, e.g., those at 0.105 m and 0.285 m, are located with antennas in the FPA w/o AS benchmark, which thus results in its much worse performance than MAs. Furthermore, although AS offers more flexibility in improving the received SNR for FPAs as compared to those w/o AS, the positions of selected antennas still cannot achieve the optimal channel gains of MAs due to their limited location options. It is also interesting to note that the sequential update algorithm fails to update the positions of five FPAs w/ AS at 0.21 m, 0.24 m, 0.3 m, 0.33 m, and 0.36 m, thus leading to a suboptimal solution. 

\begin{figure}[!t]
\centering
\includegraphics[width=2.5in]{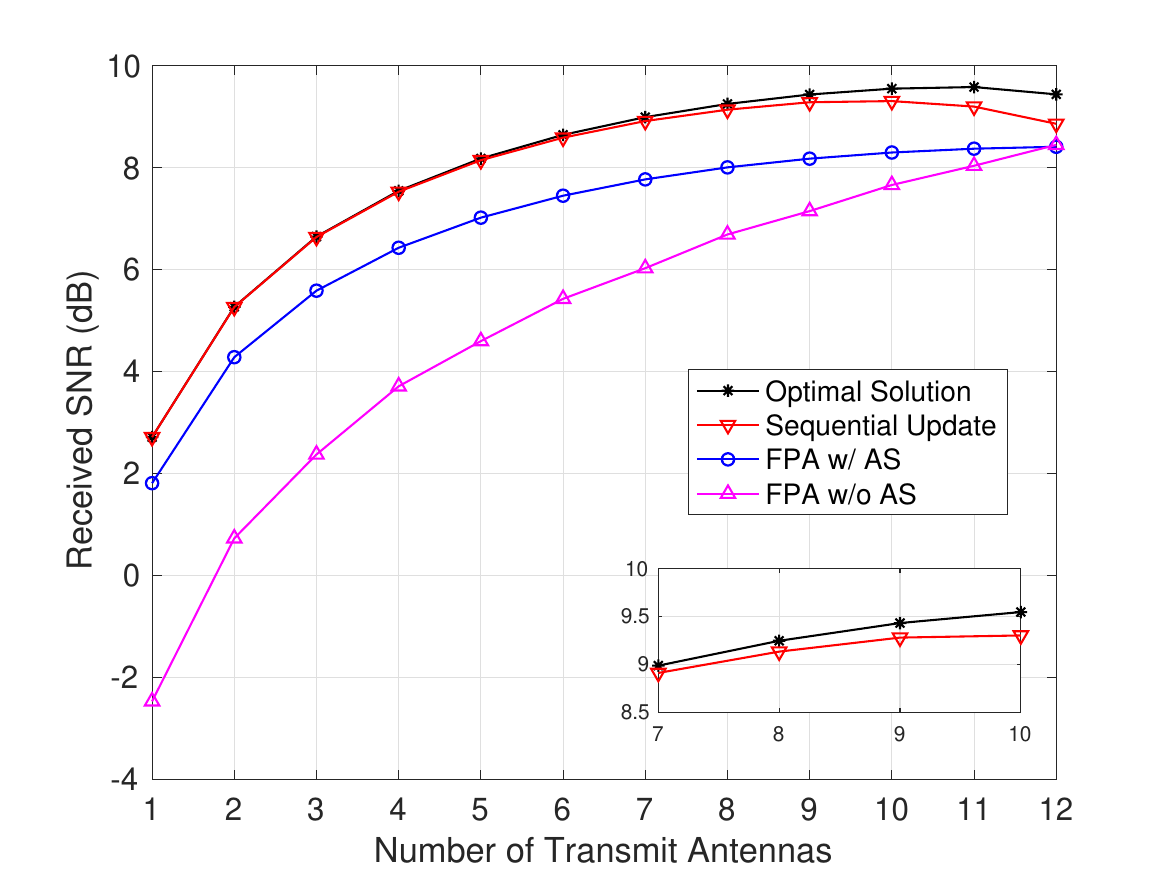}
\DeclareGraphicsExtensions.
\caption{Received SNR versus the number of MAs.}\label{SNRvsNt}
\vspace{-12pt}
\end{figure}
Next, we plot the received SNR versus the number of MAs $N$ in Fig.\,\ref{SNRvsNt}, with $M=48$. It is observed that the performance of both the proposed algorithms and the benchmark schemes improves with increasing $N$ thanks to the more significant beamforming gain. The proposed algorithms can yield considerable performance gains over the two benchmarks thanks to the flexible antenna positions. Nonetheless, such performance gains diminish as $N$ increases. This is because there exists a maximum allowable number of MAs given the distance constraints in (\ref{dmin}), i.e., $N \le L/d_{\min}=12$. Thus, when $M$ approaches 12, the flexibility of the antenna position optimization reduces, thus resulting in less performance gains. Particularly, as $N$ increases from 11 to 12, the received SNRs by the proposed algorithms even decrease due to the substantial flexibility loss, which outweighs the increased number of MAs. In addition, the sequential update algorithm is observed to achieve near-optimal performance, while its gap with the optimal algorithm increases with $N$. This is because increasing $N$ reduces the size of $\Psi_n$ in (\ref{an}) and thus limits the performance improvement in each update.

\begin{figure}[!t]
\centering
\includegraphics[width=2.5in]{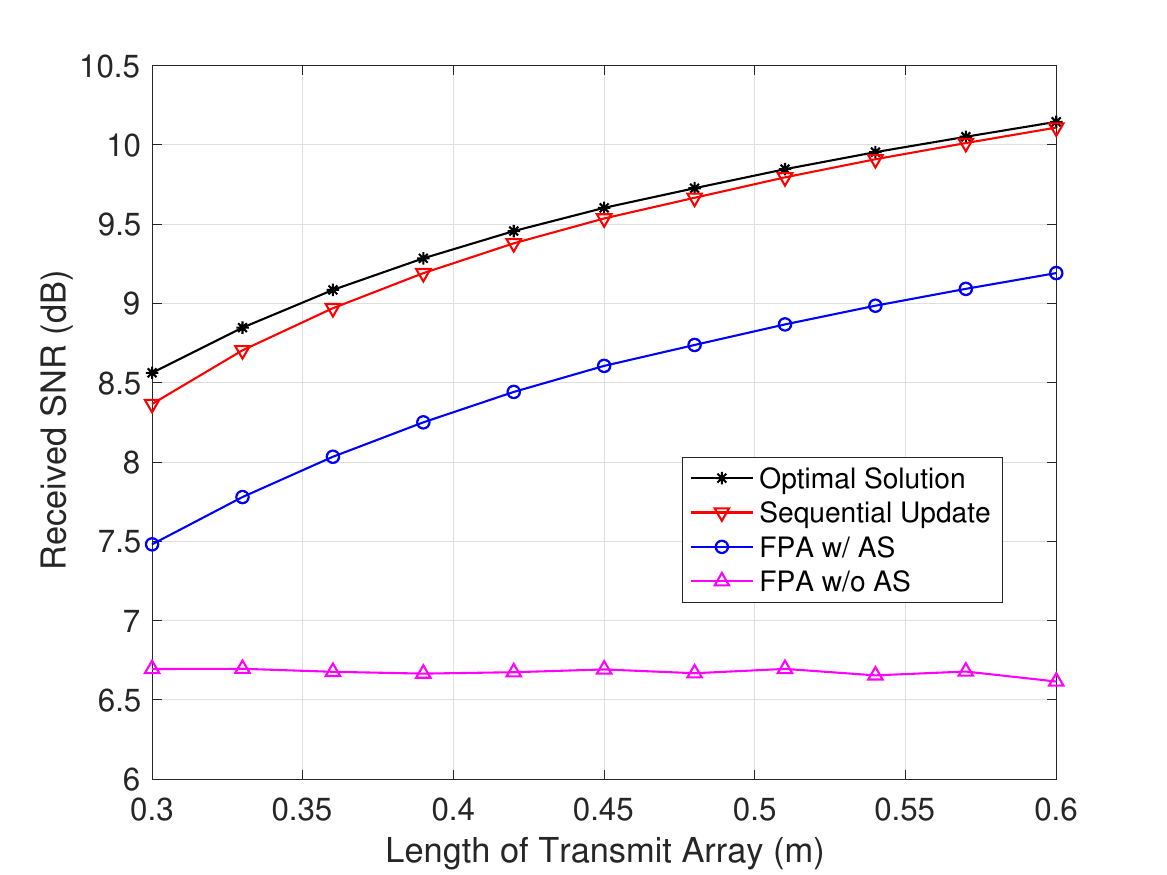}
\DeclareGraphicsExtensions.
\caption{Received SNR versus the length of transmit array.}\label{SNRvsTXSize}
\vspace{-12pt}
\end{figure}
In Fig.\,\ref{SNRvsTXSize}, we plot the received SNR versus the length of the transmit array $L$, with $N=8$ and the sampling resolution $\delta_s=0.01$ m, resulting in $M=L/\delta_s=100L$ sampling points in total. It is observed that our proposed algorithms significantly outperform the two benchmarks. In particular, the received SNRs by the proposed algorithms for MAs and the FPAs w/ AS are observed to increase with $L$ thanks to the enlarged degrees of freedom for antenna position optimization or selection. 

\begin{figure}[!t]
\centering
\includegraphics[width=2.5in]{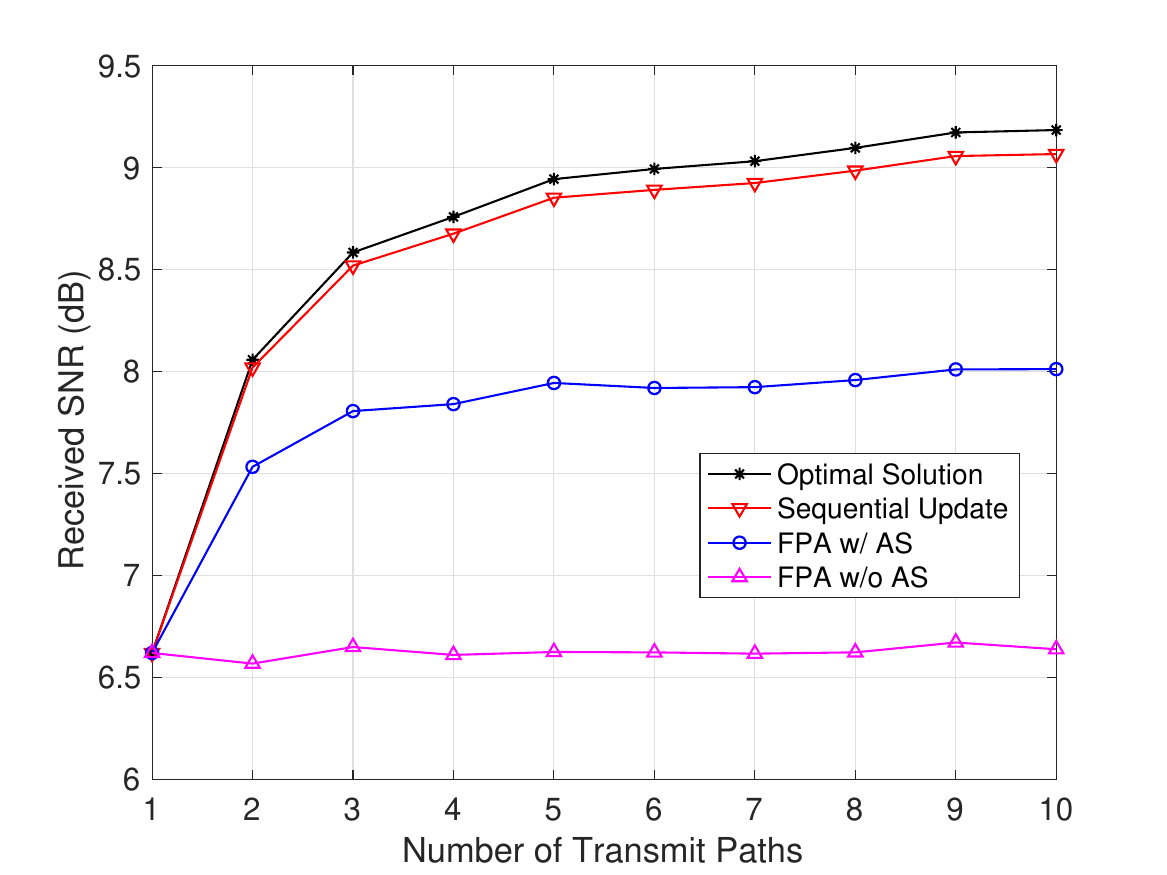}
\DeclareGraphicsExtensions.
\caption{Received SNR versus the number of transmit paths.}\label{SNRvsNumPath}
\vspace{-12pt}
\end{figure}
Finally, we plot the received SNR versus the number of transmit paths in Fig.\,\ref{SNRvsNumPath} with $M=48$ and $N=8$. It is observed that the received SNRs obtained by the proposed algorithms for MAs and the FPAs w/AS increase with the number of transmit paths, while that of the FPAs w/o AS remains almost constant. This is because small-scale multi-path channel fading becomes stronger with an increasing number of paths, such that the channel gain experiences more significant fluctuation along the transmit array, which provides performance gains by optimizing/selecting the antenna positions. However, in the case with only a single transmit path, all schemes demonstrate the same performance, since the channel gains are constant along the transmit array.\vspace{-8pt}

\section{Conclusion}
This letter considered an MA-enhanced MISO system and optimized the MA positions to maximize the received signal power. By sampling the transmit region into multiple discrete points, we proposed a graph-based algorithm to solve the point selection problem optimally in polynomial time as well as a near-optimal sequential update algorithm to solve it in linear time. Numerical results demonstrated the significant performance gains of the proposed algorithms employing MAs over the conventional FPAs with/without AS given a moderate number of sampling points for MA positions. It is interesting to extend our proposed graph-based optimization approach to MAs under more general multi-user and/or multiple-input multiple-output (MIMO) setups in the future work.\vspace{-8pt}

\bibliography{MA_new.bib}
\bibliographystyle{IEEEtran}
\end{document}